\documentclass[PRD,onecolumn,showpacs,preprintnumbers,nofootinbib,amsmath,amssymb,floatfix]{revtex4}
\usepackage{graphicx}
\usepackage{dcolumn}
\usepackage{bm}

\newcommand{\beq}{\begin{equation}}
\newcommand{\eeq}{\end{equation}}
\newcommand{\beqa}{\begin{eqnarray}}
\newcommand{\eeqa}{\end{eqnarray}}

\begin{document}

\title{Very Old Isolated Compact Objects as Dark Matter Probes}
\author{Yi-Zhong Fan}\email{yzfan@pmo.ac.cn}
\affiliation{%
\ Purple Mountain Observatory, Chinese Academy of Sciences, Nanjing
210008, China \\
\ Key Laboratory of Dark Matter and Space Astronomy, Chinese Academy of Sciences, Nanjing
210008, China}%
\author{Rui-Zhi Yang}\email{bixian85@pmo.ac.cn}
\affiliation{%
\ Purple Mountain Observatory, Chinese Academy of Sciences, Nanjing
210008, China \\
\ Key Laboratory of Dark Matter and Space Astronomy, Chinese Academy of Sciences, Nanjing
210008, China}%
\author{Jin Chang}
\email{chang@pmo.ac.cn}
\affiliation{%
\ Purple Mountain Observatory, Chinese Academy of Sciences, Nanjing
210008, China \\
\ Key Laboratory of Dark Matter and Space Astronomy, Chinese Academy of Sciences, Nanjing
210008, China}%
\begin{abstract}
Very old isolated neutron stars and white dwarfs have been suggested to be probes of dark matter. To play such a role,
 two requests should be fulfilled, i.e., the annihilation luminosity of the captured dark matter particles is above the thermal emission of the cooling compact objects (request-I) and also dominate over the energy output due to the accretion of normal matter onto the compact objects (request-II).
Request-I calls for very dense dark matter medium and the critical density sensitively depends on the residual surface temperature of the very old compact objects. The accretion of interstellar/intracluster medium onto the compact objects is governed by the physical properties of the medium and by the magnetization and rotation of the stars and may outshine the signal of dark matter annihilation.
Only in a few specific scenarios both requests are satisfied and the compact objects are dark matter burners.  The observational challenges are discussed and a possible way to identify the dark matter burners is outlined.
\end{abstract}
\maketitle

\section{Introduction}
Dark matter is a form of matter necessary to account for gravitational effects observed in very large scale structures such as the flat rotation curves of galaxies and the gravitational lensing of light by galaxy clusters that cannot be accounted for by the amount of observed/normal matter \cite{lsp,Bertone05,lkp}. In the standard cosmology model the  normal matter, (cold) dark matter,  dark energy constitute about  5\%, 23\%, 72\% of the energy density of the universe, respectively. Though abundant it is rather hard to detect the dark matter and then reveal its nature since it neither emits nor scatters photons. Among various hypothetical particles, weakly interacting massive particle (WIMP) is the leading candidate \cite{lsp,Bertone05,lkp}.

There are many experiments underway trying to catch the WIMPs both directly and indirectly \cite{Feng10}. As the Earth passes through our galaxy's dark matter halo, given the expected weak interaction scale of the WIMP-nucleon scattering, galactic WIMPs should deposit a measurable amount of energy in an appropriately sensitive detector apparatus. The measurement of the corresponding recoil energy impacted to the detector nuclei is the goal of the direct detection experiments. The indirect detection experiments, instead, look for the products of WIMP annihilations (or decays). If WIMPs are Majorana particles (for which the particles are their own antiparticles) then two colliding WIMPs would annihilate to produce gamma rays, and particle-antiparticle pairs, such as the positrons-electrons and protons-antiprotons. Another approach to the detection of WIMPs is to produce them in the laboratory, for example, with the Large Hadron Collider. So far there are some signals in favor of WIMPs but the evidence is not conclusive \cite{atic,pamela,fermi,fan,DAMA,WMAP,fermi1,CRESST,Aalseth11}.
In addition to the undergoing indirect/direct searches and accelerator experiment, the observation of stars, in particular white dwarfs and neutron stars, may also shed some light on the nature of dark matter \cite{gould12,hooper10,press85,goldman89,Bottino02,Mosk07,Bertone08,Kouvaris08,peter47,McCullough10,Kouvaris10}. The idea is the following.
The dark matter particles may be captured by the stars. The captured particles
will eventually be thermalized and centered in the core of the star, namely, a dark matter core forms in the center of
a star \cite{gould12,press85,peter47,Yang11}. The annihilation of the dark matter particles will heat the star and then give rise to some observational signals. For the very old white dwarfs and neutron stars, the annihilation of dark matter particles may be the main energy source of the surface thermal radiation and the temperature evolution will significantly deviate from the standard model in which the dark matter has not been taken into account \cite{Mosk07,Kouvaris08,peter47,McCullough10,Kouvaris10}. Consequently these two kinds of compact objects are viable probes of dark matter if two requests are fulfilled, i.e., the DM annihilation luminosity is above the thermal emission of the cooling compact object ({\it request I}) and also dominate over the energy output due to the accretion of normal matter onto the compact object ({\it request II}). For the compact objects satisfying both requests, following \cite{Mosk07} we call them the {\it dark matter burners}. In view of the uncertainties involved in the investigation, instead of carrying out an advanced numerical calculation of the dark matter capture and then annihilation in white dwarfs and neutron stars, in this work we discuss the problems analytically.

This work is structured as follows. We briefly introduce the capture of dark matter particles by white dwarfs and neutrons stars and then the annihilation of these particles in section II. The constraint on the dark matter density imposed by request-I is figured out. We then discuss the accretion of normal matter, ignored in previous literature examining the possibility of probing dark matter with neutron stars and white dwarfs, in section III. The prospects of probing dark matter with compact objects are investigated in section IV. We summarize our results with some discussion in section V.

\section{The annihilation of dark matter in very old compact objects: outshining the residual surface thermal emission}
As usual, we assume that the WIMP population has a Maxwell-Boltzmann distribution of velocities
$p(v)dv={\varrho_{\chi} \over m_\chi c^{2}} ({3\over 2\pi \bar{v}^{2}})^{3/2}4\pi v^{2}\exp(-{3v^{2}\over 2\bar{v}^{2}})d\bar{v}$,
where $\varrho_{\chi}$ ($m_\chi$) is the energy density (rest mass) of WIMPs in the neighborhood of the star, $c$ is the speed of light, and $v$ and $\bar{v}$ are the velocity and the corresponding average velocity, respectively.

The accretion rate of the WIMPs onto the neutron star with a radius $R_{_{\rm NS}}$ and a mass $M_{_{\rm NS}}$ is \cite{Kouvaris08}
\begin{equation}
F_{_{\rm NS}} \approx 2\sqrt{6\pi} {\varrho_{\chi} \over m_\chi c^{2}} \bar{v}^{-1} {GM_{_{\rm NS}}R_{_{\rm NS}}\over 1-{2GM_{_{\rm NS}}/ R_{_{\rm NS}}}}f\approx 3.5\times 10^{23}~{\rm s^{-1}}~({\varrho_{\rm \chi} \over {\rm 0.3~GeV~cm^{-3}}})({100~{\rm GeV}\over m_\chi c^{2}})
({M_{_{\rm NS}}\over 1.4M_\odot}) R_{_{\rm NS,6}} \bar{v}_{7.4}^{-1}f,
\label{eq:F_NS}
\end{equation}
where the fraction $f$ of the WIMPs that undergo one or more scatterings while inside the star and then lose enough energy to be captured can be estimated by $f=\min\{1,~\sigma_{\chi,-44.7}\}$ and $\sigma_\chi$ is the spin-independent cross section of the WIMPs scattering with the neutron/proton, and $M_\odot$ is the mass of the sun. If $\sigma_\chi$ is large enough all WIMPs passing through the star will be captured, for which we have $f=1$. In other cases we have  $f<1$. Please note that here and throughout this work, the convention $Q_{\rm x}=Q/10^{\rm x}$ has been adopted except for specific notations.

For a white dwarf with a radius $R_{_{\rm WD}}$ and a mass $M_{_{\rm WD}}$, the accretion rate is \cite{Bottino02}
\begin{equation}
F_{_{\rm WD}}\approx ({8\over 3\pi})^{1/2} {\varrho_{\chi} \over m_\chi c^{2}} \bar{v}({3v_{\rm esc}^{2}\over 2\bar{v}^{2}})\sigma_{\rm eff},
\label{eq:F_WD}
\end{equation}
where $v_{\rm esc}$ is the escape velocity of the astronomical body in question and the effective cross section is given by
$\sigma_{\rm eff}=\min[\sigma_{\rm \chi}\sum{M_{_{\rm WD}} \over m_{\rm p}}{x_{\rm i}\over A_{\rm i}}A_{\rm i}^{4}, \pi R_{_{\rm WD}}^2]$,
where $x_{\rm i}$ and $A_{\rm i}$ are the mass fraction and the atomic number of element $i$, respectively, and $m_{\rm p}$ is the rest mass of the proton.

The WIMPs scatter with nuclei and then lose their kinetic energy when
pass through a star. They will get captured if the kinetic energy is
smaller than the stellar gravity potential \cite{press85,gould12}. After being captured, the WIMPs will be
eventually thermalized in the star \cite{peter47}. The density distribution of the WIMPs is
described by \cite{nus09,ther}
\begin{equation}
\rho(r)
\approx \rho_{\rm cent} e^{-r^2/r_{\rm th}^2},
\end{equation}\\
where  $\rho_{\rm cent}$
is the number density at the center of the star and
\begin{equation}
r_{\rm th}=(\frac{3T_{\rm cent}}{2\pi m_{\rm \chi}G\rho_{\rm cent}})^{1/2}\approx 8\times 10^{9}~{\rm cm}
(\frac{T_{\rm cent}}{1\rm keV})^{1/2}(\frac{100\rm GeV}{m_{\rm \chi}c^{2}})^{1/2}\rho_{\rm cent}^{-1/2},
\end{equation}
where $T_{\rm cent}$ is the temperature at the center of the star. For the very old and ultra-cool compact objects that are of our interest, $T_{\rm cent}$ is likely controlled by the annihilation of the dark matter.
A dark matter core is formed and the annihilation efficiency is
\begin{equation}
F_{\rm A}={\int d^3 r \rho^2(r) <\sigma_{\rm A} v> \over (\int d^{3} \rho(r))^{2}}=\frac{<\sigma_{\rm A} v>}{(2\pi)^{3/2}r_{\rm th}^3},
\end{equation}
where $<\sigma_{\rm A} v> \sim 3\times 10^{-26}~{\rm cm^{3}~s^{-1}}$ is the averaged annihilation cross section.
The particle number of dark matter core evolves as
\begin{equation}
\frac{d N}{d t}=F-F_{\rm A} N^2,
\label{eq:dm_evolution}
\end{equation}
where $F$ is the capture rate (see eq.(\ref{eq:F_NS}) and eq.(\ref{eq:F_WD})) and the annihilation rate can be
expressed as $\Gamma _{\rm A} =\frac{1}{2}F_{\rm A} N^2$. Eq.(\ref{eq:dm_evolution}) can be solved analytically and one has
\begin{equation}
\Gamma_{\rm A}=\frac{1}{2}F {\rm tanh^2}(t/\tau _{\rm eq}),
\label{eq:5}
\end{equation}
where $\tau _{\rm  eq}=1/(FF_{\rm A})^{1/2}$. If $\tau _{\rm eq}$ is
longer than the lifetime of the star, the equilibrium can never be
established. However, it is straightforward to show that for the very old  compact objects considered in this work $\tau_{\rm eq}$ is much shorter than the star's lifetime $\tau_{_{\rm CO}}~(>10^{8}-10^{10}~{\rm years})$ \footnote{In the case of white dwarfs, the interior temperature is related to the surface temperature ($T_{\rm WD}$) as $T_{\rm cent}\sim 1.4\times 10^{4}~{\rm K}~(M/1~M_\odot)^{-2/7}T_{\rm WD,3}^{8/7}R_{\rm WD,9}^{4/7}$ \cite{sha87}. Consequently we have $r_{\rm th}\sim 2.7\times 10^{5}~{\rm cm}~ T_{\rm cent,4}^{1/2}(\frac{100\rm GeV}{m_{\rm \chi}})^{1/2}\rho_{\rm cent,6}^{-1/2}$ and $\tau_{\rm eq}\sim 3\times 10^{7}~{\rm s}\ll t_{_{\rm CO}}$. In the case of neutron stars with an interior temperature $\leq 2\times 10^{8}$ K, we have $T_{\rm cent}\sim 3.7\times 10^{5}~{\rm K}~(M/1~M_\odot)^{-2/7}T_{\rm NS,4}^{8/7}R_{\rm NS,6}^{4/7}$ \cite{sha87}, $r_{\rm th}\sim 100~{\rm cm}~ T_{\rm cent,5}^{1/2}(\frac{100\rm GeV}{m_{\rm \chi}})^{1/2}\rho_{\rm cent,14}^{-1/2}$ and then $\tau_{\rm eq}\sim 10^{3}~{\rm s}\ll t_{_{\rm CO}}$.}, for which the annihilation
rate is $\Gamma _{\rm A} =\frac{1}{2}F$.

For a typical neutron star, the annihilation luminosity can be estimated as
\begin{equation}
L_{_{\rm NS,ann}} \approx 5.6\times 10^{22}~{\rm erg~s^{-1}}~({\varrho_{\rm \chi} \over {\rm 0.3~GeV~cm^{-3}}})
({M_{_{\rm NS}}\over 1.4M_\odot}) R_{_{\rm NS,6}} \bar{v}_{7.4}^{-1}f,
\end{equation}
where it is assumed that most energy released in the annihilation has been deposited in the star. The corresponding surface temperature can be estimated by
\begin{equation}
T_{_{\rm NS,ann}} \sim 3000~{\rm K}~({\varrho_{\rm \chi} \over {\rm 0.3~GeV~cm^{-3}}})^{1/4} ({M_{_{\rm NS}}\over 1.4M_\odot})^{1/4} R_{_{\rm NS,6}}^{-1/4}\bar{v}_{7.4}^{-1/4}f^{1/4},
\label{eq:T_nsann}
\end{equation}
which is well consistent with the numerical result reported in Fig.1 of \cite{Kouvaris08}.
In the standard cooling model, the very old neutron star (with $\tau_{_{\rm CO}}\gtrsim 10^{8}$ years) may have a temperature as low as $T_{_{\rm NS,c}}\lesssim 10^{3}$ K \cite{Yakovlev04,Kouvaris08}. Therefore
the ``nearby" neutron stars with ${\varrho_{\rm \chi} \sim {\rm 0.3~GeV~cm^{-3}}}$, in principle can be a viable probe of dark matter. The caution is that the magnetic field decay of some highly magnetized neutron stars may heat the neutron star significantly \cite{Augi08}. One should also bear in mind that it is lack of observational evidence for old neutron stars cooler than $\sim 10^{4}$ K (The coolest neutron star detected so far has an effective temperature $\sim 8\times 10^{4}$ K \cite{Kargaltsev04}). Therefore we suggest a ``conservative" residual surface temperature of the very old neutron star $T_{_{\rm NS,c}}\sim 10^{4}$ K, with which a viable dark matter probe calls for a dense dark matter medium
\begin{equation}
\varrho_{\rm \chi} \gtrsim  \varrho^{\rm I}_{\rm NS,\chi-c} \sim 38~{\rm GeV~cm^{-3}}~({T_{_{\rm NS,c}}\over 10^{4}~{\rm K}})^{4}R_{_{\rm NS,6}}({M_{_{\rm NS}}\over 1.4M_\odot})^{-1}f^{-1}\bar{v}_{7.4}.
\label{eq:NS}
\end{equation}
Since $\varrho^{\rm I}_{\rm NS,\chi-c}$ is much higher than the energy density of dark matter in most regions of our Galaxy, the lack of neutron stars cooler than $10^{4}$ K is unlikely due to dark matter heating.

The luminosity of the white dwarf with a mass $\sim 0.5~M_\odot$ and a radius $\sim 10^{9}$ cm purely due to the accretion and subsequent annihilation of DM is given by \cite{Bertone08}
\begin{eqnarray}
L_{_{\rm WD,ann}} \approx  2\times 10^{22}~{\rm erg~s^{-1}}~({\varrho_{\rm \chi} \over 0.3~{\rm GeV~cm^{-3}}})({M_{_{\rm WD}}\over 0.5M_\odot})^{2}
\sigma_{\rm \chi,-44} \bar{v}_{7.4}^{-1}R_{_{\rm WD,9}}^{-1}.
\end{eqnarray}
For parameters $\varrho_{\rm \chi} \sim 10^{10}~{\rm GeV~cm^{-3}}$, $m_\chi c^2 \sim 100~{\rm GeV}$, $\sigma_{\rm chi,-44}\sim 10$, $M_{_{\rm WD}}\sim 0.6~M_\odot$
and $\bar{v}_{7.4}\sim 1$, we have $F_{_{\rm WD,ann}}\sim 6\times 10^{34}~{\rm s^{-1}}$, which is smaller than the corresponding numerical result reported in Fig.1 of \cite{Mosk07} by a factor of $\sim 2$, accurate enough for our purpose.

The surface temperature is
\begin{eqnarray}
T_{_{\rm WD,ann}} \sim  72~{\rm K}~({\varrho_{\rm \chi} \over {\rm 0.3~GeV~cm^{-3}}})^{1/4}R_{_{\rm WD,9}}^{-3/4}
({M_{_{\rm WD}}\over 0.5M_\odot})^{1/2}\sigma_{\rm \chi,-44}^{1/4}\bar{v}_{7.4}^{-1/4}.
\label{eq:T_wdann}
\end{eqnarray}
Such a temperature is considerably lower than the expected residual temperature $T_{_{\rm WD,c}}\gtrsim 10^{3}~{\rm K}$ of the ultra-cool white dwarf\footnote{The coolest white dwarf detected so far has an effective temperature about 3500 K and theoretically the coolest white dwarfs may have an effective temperature as low as $\sim 2000$ K. The project aiming to detecting such ultra-cool white dwarfs via combing deep infrared and optical data is ongoing \cite{Catalan10}.} with a lifetime as long as the Universe \cite{Chen11}. To be a viable dark matter probe, the white dwarfs should be surrounded by dark matter with an energy density
\begin{equation}
\varrho_{\rm \chi} \gtrsim \varrho^{\rm I}_{\rm WD,\chi-c} \sim 10~{\rm TeV~cm^{-3}}~({T_{_{\rm WD,c}}\over 10^{3}~{\rm K}})^{4}R_{_{\rm WD,9}}^{3}({M_{_{\rm WD}}\over 0.5M_\odot})^{-2}\sigma_{\rm \chi,-44}^{-1}\bar{v}_{7.4}.
\label{eq:WD}
\end{equation}


\section{Accretion of normal matter: one potential challenge of probing dark matter with compact objects}
The accretion of normal matter, either the interstellar/intracluster medium or the material ejected from a binary star, onto the compact objects is quite common in nature. A fraction of the gravitational energy of the accreting material will be released on the surface of the compact objects and then give rise to strong thermal radiation, which may outshine the dark matter annihilation signal and then renders the identification of dark matter burner more challenging. This is in particular the case for the compact objects in the binary system, for which very strong emission has been well detected \cite{sha87}. That's why in this work we only discuss the very old isolated compact objects.

\subsection{The un-magnetization case}
Here we discuss the simplest scenario in which the compact objects are un-magnetized (or very weakly magnetized).
In such a case, the mass flux of the accretion onto a point mass, for example, an isolated neutron star or white dwarf,  can be estimated as \cite{Bondi52}
\begin{eqnarray}
\dot{M}={4\pi \lambda_{\rm s} G^{2}M^{2}\varrho_\infty \over (v^{2}_\infty+V^{2})^{3/2}c^{2}}
\sim 10^{8}~{\rm g~s^{-1}}~\lambda_{\rm s}({M\over 1~M_\odot})^{2}({\varrho_\infty\over 1.6~\rm GeV~ cm^{-3}})({\bar{V}\over 10^{7}~{\rm cm/s}})^{-3},
\label{eq:Bondi}
\end{eqnarray}
where the parameter $\lambda_{\rm s}$ ranges from 0.25 for $v_\infty\geq V$ to unity for $v_\infty\ll V$, $\varrho_\infty$ is the energy density of the surrounding medium, $\bar{V}\equiv \sqrt{v^{2}_\infty+V^{2}}$, $v_\infty=\sqrt{kT_\infty/m_{\rm p}}$, and $V$ is the velocity of the compact object moving relative to an ambient medium.

The total luminosity due to the accretion onto a neutron star is estimated to be
\begin{eqnarray}
L_{_{\rm NS,acc}}
\sim 3.7\times 10^{28}~{\rm erg~s^{-1}}~({M_{_{\rm NS}}\over 1.4~M_\odot})^{3}({\varrho_\infty\over 1.6~\rm GeV~cm^{-3}})
\bar{V}_{7}^{-3} R_{_{\rm NS,6}}^{-1},
\end{eqnarray}
where it is assumed that $V\gg v_\infty$.

For white dwarf we have
\begin{eqnarray}
L_{_{\rm WD,acc}}
\sim 4.5\times 10^{26}~{\rm erg~s^{-1}}~({M_{_{\rm WD}}\over 0.5~M_\odot})^{3}({\varrho_\infty\over 1.6~\rm GeV~ cm^{-3}})
T_{\infty,4}^{-3/2} R_{_{\rm WD,9}}^{-1},
\end{eqnarray}
where it is assumed that $V\leq v_\infty$.

Suppose the main energy source of the surface emission of white dwarfs is the annihilation of dark matter particles, it is required that
\begin{eqnarray}
\varrho_{\rm \chi} \geq \varrho^{\rm II}_{\rm WD,\chi-c} \sim 6.7~{\rm TeV~cm^{-3}}~({M_{_{\rm WD}}\over 0.5~M_\odot})({\varrho_\infty\over 1.6~\rm GeV~ cm^{-3}})
T_{\infty,4}^{-3/2} \sigma_{\rm \chi,-44}^{-1}\bar{v}_{7.4}.
\label{eq:WD-1}
\end{eqnarray}

For neutron stars such a request changes to be
\begin{eqnarray}
\varrho_{\rm \chi} \geq  \varrho^{\rm II}_{\rm NS,\chi-c} \sim 200~{\rm TeV~cm^{-3}}~({M_{_{\rm NS}}\over 1.4~M_\odot})^{2}R_{\rm NS,6}^{-2}({\varrho_\infty\over 1.6~\rm GeV~cm^{-3}}) \bar{V}_{7}^{-3}f^{-1}\bar{v}_{7.4}.
\label{eq:Requst-NS-II}
\end{eqnarray}

Obviously the required dark matter energy density is very high unless the interstellar/intracluster medium surrounding the white dwarf is extremely hot (e.g., $T_{\infty}\geq 10^{6}$ K) or the neutron star has a very high kick velocity (e.g., $V>10^{8}~{\rm cm~s^{-1}}$). At a distances $\sim 100$ pc to the Earth, interstellar absorption measurements suggest that the solar system is embedded in an anomalously low-density region.  This is widely believed to contain hot ($\sim 10^{6}$ K), low-density ($\sim 5\times 10^{-3}~{\rm cm^{-3}}$) plasma giving rise to the extreme ultraviolet/soft X-ray background \cite{Paresce84}. Eq.(\ref{eq:WD-1}) thus gives
\[
\varrho^{\rm II}_{\rm WD,\chi-c} \sim 0.03~{\rm GeV~cm^{-3}}~({M_{_{\rm WD}}\over 0.5~M_\odot})({\varrho_\infty\over 8\times 10^{-3}~\rm GeV~cm^{-3}}) T_{\infty,6}^{-3/2} \sigma_{\rm \chi,-44}^{-1}\bar{v}_{7.4},
\]
which is {\it below} the widely-accepted energy density $\sim 0.3~{\rm GeV~cm^{-3}}$ of the dark matter at the same site. Therefore for the old white dwarfs isolated in such a hot low-density cavity, the energy released per time via accretion of normal matter is lower than that via the annihilation of the captured dark matter (i.e., request-II is fulfilled). For neutron stars, since $\bar{V}$ is still in order of $10^{7}~{\rm cm~s^{-1}}$ the decrease of $\varrho_\infty$ by a factor of $\sim 200$ is not enough to meet request-II.

\subsection{The magnetization case}
In the case of un-magnetization, the probe of dark matter with compact objects seems a bit challenging. Fortunately
 both neutron stars and white dwarfs are likely magnetized with a field strength $B_*$ and the magnetization may play a key role in suppressing the accretion onto some compact objects, rendering the probe of dark matter with compact objects, in particular neutron stars, more promisingly.

Let's consider a compact object isolated in the middle of a uniform gaseous medium. The accreting matter will fall in radially at large distances, where the magnetic pressure is small. The infall will be considerably deflected from a radial flow at a radius $r_{\rm A}$ (i.e., the so-called Alfven radius) for a non-rotating magnetized compact object, where the magnetic energy density is equal to the kinetic energy density of the gas. The continuity equation reads  $\rho_{\rm gas}
 u=-\dot{M}/4\pi r^{2}$ and the infall velocity of an ionized gas inside the accretion radius can be estimated by
$u=-({GM/2r})^{1/2}$, where $\rho_{\rm gas}$ is the mass density of the inflowing material.
The kinetic energy density is then estimated as
\begin{equation}
e_{\rm kin}={1\over 2}\rho_{\rm gas} u^{2}\approx {\dot{M}\sqrt{GM}\over 8\pi \sqrt{2}}r^{-5/2}.
\end{equation}
Note that at $r\gg R_*$, the amplitude of dipole magnetic field is $B \approx \sqrt{2}\mu/r^{3}$, where $\mu=B_* R_*^{3}$ is the dipole moment, and $R_*$ is the radius of the compact object. The magnetic energy density can be estimated by $e_{\rm mag} \approx \mu^{2}/4\pi r^{6}$ and consequently
the Alfven radius can be estimated by \cite{Blaes93,Treves00}
\begin{equation}
r_{_{\rm A}}\approx 3.6\times 10^{10}~{\rm cm}~\mu_{30}^{4/7}\dot{M}_{10}^{-2/7}(M/1~M_\odot)^{-1/7}.
\end{equation}

For the compact objects with $R_* \lesssim r_{_{\rm A}}$, the accreted material may be compelled to flow along the magnetic field lines. To estimate whether significant accretion can take place or not, the so-called accretion radius and corotation radius are relevant.
The accretion radius defines the region where the dynamics of the interstellar/intracluster medium are dominated by the gravitational field of the compact object and is estimated by \cite{Bondi52}
\begin{equation}
r_{\rm acc}\sim {GM \over V^{2}+c_\infty^{2}}\sim 1.4\times 10^{12}~{\rm cm}~(M/1~M_\odot)\bar{V}_{7}^{-2}.
\end{equation}
The corotation radius at which the angular velocity of the compact object equals to the Keplerian angular velocity is \cite{Treves00}
\begin{equation}
r_{\rm co} \sim 2 \times 10^{8}~{\rm cm}~(M/1~M_\odot)^{1/2}P^{2/3},
\end{equation}
where $P$ is the spin period.

First, the accretion is suppressed for $r_{_{\rm A}}>r_{\rm acc}$, i.e.,
\begin{equation}
\dot{M} \leq M_{\rm crit}=2.7\times 10^{4}~{\rm g~s^{-1}}~\mu_{30}^{2}({M \over 1M_\odot})^{-4}\bar{V}_{7}^{7},
\end{equation}
in which case the system remains in the so-called {\it georotator} stage. With eq.(\ref{eq:Bondi}), the above request equals to
\begin{equation}
\mu>\mu_{\rm crit}\simeq 6.4\times 10^{31}~{\rm Gauss~cm^{3}}~\lambda_{\rm s}^{1/2}(M/1~M_\odot)^{3}\bar{V}_{7}^{-5}(\varrho_\infty/1.6~{\rm GeV~cm^{-3}})^{1/2}.
\label{eq:mu_crit}
\end{equation}

Second, the accretion is suppressed if at $r_{\rm acc}$ the gravitational energy density of the inflowing material ($e_{_{\rm G}} \sim 6.5\times 10^{-9}~{\rm erg~cm^{-3}}~\dot{M}_{10}r_{\rm acc,12}^{-5/2}(M/1~M_\odot)^{1/2}$) is smaller than the energy density of the relativistic momentum outflow produced by the rotating magnetic field (for a dipole field, $e_{_{\rm B}} \sim 7.5\times 10^{-5}~{\rm erg~cm^{-3}}~B_{12}^{2}P^{-4}R_{6}^{6}r_{\rm acc,12}^{-2}$), requiring \cite{Blaes93}
\begin{equation}
P\lesssim P_{\rm crit}\simeq 30~{\rm s}~\mu_{30}^{1/2}\lambda_{\rm s}^{-1/4}(M/1M_\odot)^{-1/2}({\varrho_\infty\over 1.6~{\rm GeV~cm^{-3}}})^{-1/4} \bar{V}_{7}^{1/2}.
\label{eq:P_crit}
\end{equation}
Such a system is in the so-called {\it ejector} phase and there is no accretion (For an initial rotation with a period longer than $P_{\rm crit}$, the below discussion is irrelevant).
Suppose the gravitational wave radiation can be ignored and the star is slowing down at the magnetic dipole rate, $P\gtrsim P_{\rm crit}$ is reached at a time
\begin{equation}
t_{_{\rm B}} \sim 29~{\rm Gyr}~ \mu_{30}^{-1}\lambda_{\rm s}^{-1/2}\bar{V}_{7}(M/1M_\odot)^{-1}(\varrho_\infty/1.6~{\rm GeV~cm^{-3}})^{-1/2}I_{45},
\label{eq:t_b}
\end{equation}
where $I$ is the typical moment of inertia of the compact object. As long as $t_{_{\rm CO}}<t_{_{\rm B}}$,  eq.(\ref{eq:P_crit}) holds.

\begin{figure}
\includegraphics[width=150mm,angle=0]{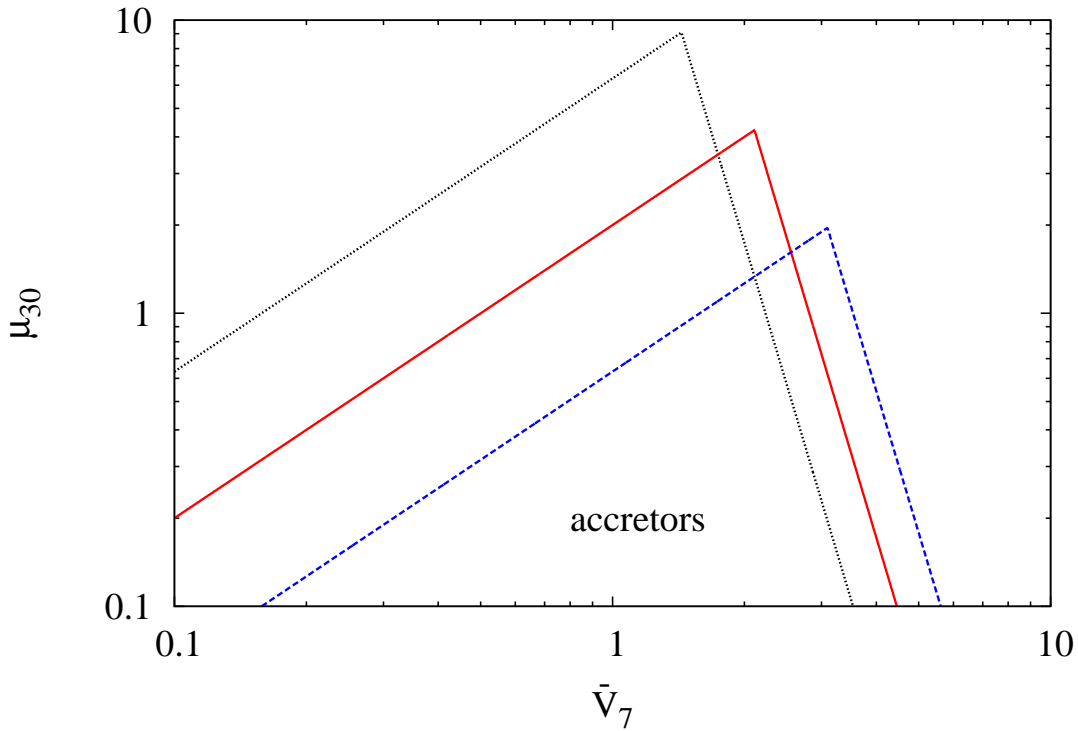}
  \caption{The strong magnetization and fast rotation of an old isolated neutron star may hamper the accretion of the interstellar/intrucluster medium. The dotted, solid and dashed lines represent $\varrho_{\rm \infty}\sim (0.16,~1.6,~16)~{\rm GeV~cm^{-3}}$, respectively. In the regions above the lines, no accretion takes place. Please note that $M_{_{\rm NS}} \sim 1.4M_\odot$, $I_{45}\sim 1$, $R_{_{\rm NS}}\sim 10^{6}$ cm, $\lambda_{\rm s}\sim 1$, and $t_{_{\rm CO}}\sim 10~{\rm Gy}$ have been taken into account.}
  \label{fig:1}
\end{figure}

After $P$ has increased above $P_{\rm crit}$, the infalling material proceeds undistributed until the Alfv\'en radius. The co-rotating magnetosphere will then prevent the accreting material from going any further if $r_{_{\rm A}}\gtrsim r_{\rm co}$, i.e.,
\begin{equation}
P\lesssim P_{_{\rm A}} \sim 1.7\times 10^{4}~{\rm s}~\mu_{30}^{6/7}(\varrho_\infty/1.6~{\rm GeV~cm^{-3}})^{-3/7}\bar{V}_{7}^{9/7}(M/1~M_\odot)^{-51/28}.
\label{eq:P_A}
\end{equation}
Such a system remains in the so-called {\it propeller} phase.
For most neutron stars, $P_{_{\rm A}}$ is much longer than $P_{\rm crit}$ and $P$, preventing the accretion. The accreted material, however, will be spun up by the magnetosphere and therefore exert a torque of its own on the neutron star. A full two-dimensional or even three-dimensional MHD investigation of the interaction between accretion flow and rotating magnetosphere may be needed to understand such a physical process thoroughly. An approximated expression for the the torque was presented in \cite{Blaes93}, with which the corresponding spin-down timescale is
$t\sim 5~{\rm Gyr}~B_{12}^{-11/14}(\varrho_\infty/1.6~{\rm GeV~cm^{-3}})^{-17/28}\bar{V}_{7}^{2}$,
allowing interstellar accretion. The other kind of possibility is that although the interchange instabilities of the magnetospheric boundary during the subsonic propeller state are suppressed, the ``magnetic gates" are not closed completely since the atmospheric plasma may be able to penetrate into the stellar magnetic field due to diffusion. The diffusion rate is limited to
$\dot{M}_{\rm dif} \lesssim 1.2\times 10^{7}~{\rm g~s^{-1}}~\zeta_{-1}^{1\over 2}\mu_{30}^{-{1\over 14}}({M\over 1M_\odot})^{1\over 7}\dot{M}_{10}^{11\over 14}$,
where $\zeta \sim 0.1$ is the diffusion efficiency \cite{Ikhsanov01}.
In view of these possibilities, following \cite{Blaes93,Treves00} we assume that the barrier given by eq.(\ref{eq:P_A}) can be overcome for a considerable fraction of neutron stars. Hence to prevent the accretion, either the  $P_{\rm crit}$ constraint or the $\mu_{\rm crit}$ constraint must be satisfied. Please see Fig.\ref{fig:1} for the regions in which the accretion is hampered.

Most white dwarfs rotate slowly, with a period hours even days \cite{sha87}. For these old white dwarfs the $P_{\rm crit}$ constraint or even the $P_{_{\rm A}}$ constraint is violated. The $\mu_{\rm crit}$ constraint is also violated  unless the white dwarfs are in the very hot low density medium with a $T_\infty > 10^{6}~{\rm K}$. Therefore these white dwarfs are normal-matter accretors and the magnetization does not help in probing the nature of dark matter.
Most neutron stars detected so far, instead, rotate very quickly ($P<10~{\rm s}$) and have a typical kick velocity $V\sim 3\times~ 10^{7}~{\rm cm~s^{-1}}$ \cite{Arzoumanian02}, with also a $B_{*}\geq 10^{12}$ Gauss \cite{sha87}, the $P_{\rm crit}$ and/or $\mu_{\rm crit}$ constraints are satisfied, suggesting that for a considerable fraction of neutron stars the accretion of normal matter has been hampered thanks to their fast rotation and strong magnetization. To probe dark matter with these neutron stars  just request-I (i.e., eq.(\ref{eq:NS})) should be taken into account.

\section{The prospect of probing dark matter with compact objects}
The number densities of local white dwarfs and neutron stars are estimated to be $1.5\times 10^7~{\rm kpc^{-3}}$ and
$2\times 10^6~{\rm{kpc^{-3}}}$, respectively \cite{sha87}. Some isolated white dwarfs are found to be very cool and are almost as old as the Universe.
There should be some very old and cool isolated neutron stars, which may be a bit under-luminous to be detected by current telescopes, in the Galaxy.

As shown in section III.B, the magnetization of most white dwarfs won't hamper the accretion of normal matter. Moreover, the dipole radiation of a typical white dwarf is
$L_{\rm WD,dip} \sim 6\times 10^{19}~{\rm erg~s^{-1}}~B_{\rm dip,3}^{2}R_{\rm WD,9}^{6}P_{\rm WD,3}^{-4}$,
where $B_{\rm dip}\sim B_*$ is the strength of the dipole magnetic field. Such a luminosity is too low to be comparable to the dark matter annihilation signal that is of our interest. Therefore white dwarfs are a viable dark matter probe as long as the request that (see eq.(\ref{eq:WD}) and eq.(\ref{eq:WD-1}))
\begin{equation}
\varrho_{\rm \chi}\geq \max\{\varrho^{\rm I}_{\rm WD,\chi-c},~\varrho^{\rm II}_{\rm WD,\chi-c}\}
\label{eq:WD-Basic}
\end{equation}
is satisfied.

For a very old neutron star, the dipole radiation luminosity can be estimated by
$L_{\rm dip} \sim 5.6\times 10^{27}~{\rm erg~s^{-1}}~B_{\rm dip,12}^{2}R_{\rm NS,9}^{6}P_{\rm NS,1}^{-4}$. Such a luminosity is above the expected dark matter annihilation luminosity $(\sim 1.9\times 10^{26}~{\rm erg~s^{-1}}(\varrho_{\rm \chi}/1~{\rm TeV~cm^{-3}})\bar{v}_{7.4}^{-1}f)$ of neutron stars for reasonable parameters. Fortunately only a small fraction of the dipole luminosity will be re-radiated in the optical band, so the dark matter annihilation signal may be distinguished. The other possibility is that for the very old neutron stars, its surface magnetic field might have decayed significantly and consequently the dipole radiation could be much weaker. Therefore neutron stars can be a viable dark matter probe as long as the request  that
$\varrho_{\rm \chi} \gtrsim  \varrho^{\rm I}_{\rm NS,\chi-c}$
(i.e., eq.(\ref{eq:NS}))
is satisfied, where we have taken into account the fact that for most neutron stars the accretion of normal matter has been hampered (i.e., eq.(\ref{eq:P_crit})).

For our purpose the white dwarfs or neutron stars should be surrounded by very dense dark matter. Such a request is somewhat hard to fulfil. The center of our Galaxy is expected to be dark-matter-dense. Indeed most works on probing dark matter with compact objects focus on the very inner ($\lesssim 1$ pc) region of the Galactic center \cite[e.g.,][]{goldman89,Mosk07,McCullough10}. Because of interstellar dust along the line of sight, the Galactic Center cannot be studied at visible, ultraviolet or soft X-ray wavelengths. The available information about the Galactic center comes from observations at gamma ray, hard X-ray, infrared, sub-millimeter and radio wavelengths. If $T_{_{\rm NS,c}}$ is indeed $\sim {\rm a~few}\times 10^{4}$ K, it is almost impossible to detect these neutron stars mainly emitting optical/ultraviolet photons. The situation is somewhat different for white dwarfs since the dark-matter-heating may have a signal in infrared band, which suffers from much weaker dust extinction. However the Galactic center is so bright in infrared band \cite{Schlegel98} that the identification of a very cool white dwarf is not an easy task. In addition to our Galactic center, some Galactic global clusters might be dark-matter-rich \cite{Mash05}. Our Galaxy has about 250 globular clusters, all of them in highly elongated orbits. On the one hand, the globular clusters lying outside the disk do not suffer from significant dust extinction. On the other hand, the globular clusters are likely gas and dust free. For example, the intracluster medium of the global clusters 47 Tuc, M2, M15 and NGC 6624 has been detected to be $\sim (0.1,~3,~0.3,~0.01)~M_\odot$, respectively. The typical energy density of intracluster medium is then $ \sim 0.16~{\rm GeV~cm^{-3}}$ \cite{Moore11}. It is speculated in \cite{Bertone08} that in the center ($\leq 1~{\rm pc}$) of the global cluster M4  the velocity of the dark matter is $\sim 20~{\rm km/s}$ and the dark matter density might be $\gtrsim 10^{2}~M_\odot~{\rm pc}^{-3} \sim 4~{\rm TeV~cm^{-3}}$. Such a dark matter medium is likely dense enough to make the inhabited white dwarfs  the dark matter burners (it is straightforward to show that eq.(\ref{eq:WD-Basic}) is satisfied for current parameters) as long as $T_{_{\rm WD,c}}< 3\times 10^{3}$ K. The same holds for the neutron stars. Therefore the center of some Galactic globular clusters, rather than that of our Galaxy, may be a good place to search for dark matter burners. There is however one caution. So far it is still lack of solid observational evidence for the existence of a significant dark matter component in Galactic globular clusters \cite{Lane10,Conroy11}. The dwarf spheroidal galaxies are characterized by little to no gas or dust or recent star formation. Though very faint, these galaxies are likely the most dark matter-dominated galaxies. For example, Ursa Major II, at a distance $\sim 32$ kpc, has an extremely low luminosity $\sim 2.8\times 10^{3}L_\odot$ (where $L_\odot$ is the luminosity of the sun) but a rather high total mass $\sim 3.1^{+5.6}_{-1.8}\times 10^{6}~M_\odot$ within 100 pc \cite{Strigari08}. The very inner region of such galaxies may be so dark matter-dense that can host a few dark matter burners. Given the very low dust extinction and/or gas absorption, the detection of these dark matter burners in principle is possible. One challenge is that these objects may be too far/faint to be identified.

\section{Conclusion and Discussion}
In this work, we have analytically discussed the dark matter annihilation in white dwarfs and neutron stars and the prospects of probing dark matter with these two kinds of compact objects. To be a viable dark matter probe, two requests should be satisfied, including that the luminosity arising from the annihilation of the captured dark matter particles is above the thermal luminosity of the cooling compact object (request-I) and also dominate over the energy output due to the accretion of normal matter onto the compact object (request-II). The first request has been noticed before while the second request has been ignored in previous literature. We show in section III that the accretion of normal matter onto the compact objects may play an important role in heating the stars.
 The strong magnetization and the fast rotation of a considerable fraction of neutron stars can hamper the accretion of the normal matter. For most white dwarfs the weak magnetization and the slow rotation are unable to prevent the accretion of normal matter. Nevertheless, to be a viable probe, the dark matter medium surrounding the compact objects should be very dense. The suitable sites are very limited. It is widely believed that our Galactic center is dark-matter-rich. However, our Galactic center is also rich of gas and dust, which renders the detection in optical/ultraviolet band rather challenging or even impossible. Consequently the detection of dark-matter-heating neutron stars with a surface thermal emission plausibly in optical/ultraviolet band is unlikely. The dark-matter-heating old white dwarfs likely peak in infrared band, for which the detection prospect is relatively better. However, our Galactic center is very bright in infrared band and the identification of one single ultra-cool white dwarf is not an easy job. The center of some Galactic globular clusters, if indeed dark-matter-dense, will be a better place to search for dark-matter-heating compact objects thanks to the low density of gas and dust and the lying outside the Galactic disk. This possibility will be much more attractive if unambiguous observational evidence for the existence of a significant dark matter component in Galactic globular clusters can be obtained in the near future. The dwarf spheroidal galaxies are likely the most dark matter-dominated galaxies and are dust/gas free. Their center may host some dark matter burners, which, however, may be too far/faint to be identified.

Even we have got some candidates, to distinguish between the dark matter burners and the cooling compact objects, both can give rise to a prominent thermal emission signal, is non-trivial. One possible solution is the following. The dark matter density is expected to drop with the radius to the center of the Galaxy or the Galactic globular cluster or the dwarf spheroidal galaxiy, so is the dark matter annihilation luminosity (i.e., the outer the compact objects, the lower the dimmest surface radiation luminosity of the neutron stars or white dwarfs). If such a tendency can be found in the future, the dark matter burner nature will be favored.

\acknowledgments We are grateful to Bing Zhang and Daming Wei for the discussion on the accretion of normal matter onto the magnetized compact objects, and Dong Xu and Jingzhi Yan for the discussion on the detectability of dark matter burners at various sites. We also thank the anonymous referee for helpful suggestions. This work was supported in part by National Basic Research Program of China under grants 2009CB824800 and 2010CB0032, and by National Natural Science of Chinese under grants 10920101070, 10925315 and 11073057. YZF is also supported by the 100 Talents Program of Chinese Academy of Sciences.

\end{document}